\documentclass{aa}
\usepackage{txfonts}

\usepackage[dvips]{graphicx}
\usepackage[]{longtable}

\def\beq{\begin{equation}}
\def\eeq{\end{equation}}

\def\beq{\begin{equation}}
\def\eeq{\end{equation}}
\def\et{\textit{et al.~}}
\def\teff{$T_\mathrm{eff}$}
\def\rstar{$R_\ast$}
\def\rsun{$R_\odot$}
\def\rjup{$R_{\mathrm{Jup}}$}

\begin{document}

\title{Characterization of extrasolar planetary transiting candidates II
       }

\author{
J. Gallardo \inst{1}, 
S. Silva \inst{1}, 
S. Ram\'{\i}rez Alegr\'{\i}a \inst{2,3},
D. Minniti \inst{2,4}
and
P. Pietrukowicz \inst{2,5}
}

\offprints{J. Gallardo, \email{gallardo@das.uchile.cl}}

\institute{Departamento de Astronom\'{\i}a, Universidad de Chile, Casilla 36-D, Santiago, Chile
     \and
Departamento de Astronom\'{\i}a y Astrof\'{\i}sica, P. Universidad Cat\'olica de Chile
\and
Instituto Astrof\'{\i}sica de Canarias, 38200, La Laguna, Tenerife, Spain 
     \and
Vatican Observatory, V00120 Vatican City State, Italy
     \and 
Nicolaus Copernicus Astronomical Center, ul. Bartycka 18, 00-716 Warszawa, Poland
          }

\date{Received ... / Accepted ... }

\titlerunning{Extrasolar planetary transit candidates}
\authorrunning{Gallardo et al.}

\abstract{
We present a second paper of fully characterization of a sample of stars whose low-depth transits
were discovered by the OGLE-III campaign in order to select the most
promising candidates for spectroscopic confirmation, following the same analysis done in Gallardo \et (2005). We present new optical and near-IR
photometry, deriving physical parameters like effective
temperature (\teff), distance ($d$), the stellar radii (\rstar) and
the companion radii ($R_\mathrm{c}$). We selected eight M (2800 K
$\leq$~\teff~$\leq$ 3850 K) or K (3850 K
$\leq$~\teff~$\leq$ 5150 K) spectral type 
stellar objects as potential candidates to host exoplanets, even
though, considering the radii of their companions, only stars
OGLE-TR-61, OGLE-TR-74, OGLE-TR-123 and
OGLE-TR-173 are the most promising M-type transit candidates to host
planets. Confirmation of the planetary nature of any of these objects will yield another transiting extrasolar planet orbiting a
M-type star, or even more interesting, the first extrasolar planets orbiting a late M-type dwarf of effective temperature about 2900 K.

}%

\maketitle

\section{Introduction}

The first transiting extrasolar planet was announced in 2000, when Charbonneau \et (2000) and Henry \et (2000) confirmed, via radial velocity measurements, the planetary nature of the transit light curve of HD 209458b. As of February 2010, there are approximately 69 extrasolar planets who transit their parent stars. Many of these planetary companions have been discovered by different photometric extra-solar planets surveys such as OGLE (Optical Gravitational Lensing Experiment), STARE (STellar Astrophysics \& Research on Exoplanets), WASP (Wide Angle Search for Planets) and very recently by the COROT space telescope (COnvection, ROtation \& planetary Transits, lunched at the end of 2006), which claimed out the discovered of the first telluric planet (L\'eger \et (2009)). In despite of the success of these projects, one of the major problems identifying transiting planets is that photometry alone does not allow to unambiguously distinguish between extrasolar planets and other astrophysical low-luminosity objects like very late M-type stars, brown dwarfs or grazing binaries, as all these objects have similar characteristics (e.g., Dreizler \et 2002, 2007; Gallardo \et 2005; Silva \& Cruz 2006). It is evident that transit surveys may monitor orders of magnitude more stars in a comparatively shorter observing time than radial velocity surveys, and hence they have the ability to discover many interesting exoplanet candidates, but non-planetary events must be filtered out. Once the classification is made, precise spectroscopic observations enable us to measure all the important parameters such as radii and masses.  

Several works present efficient ways to select the most promising planetary candidates from transit surveys. For instance, Dreizler \et (2002, 2007) made a selection using low dispersion spectroscopy; Drake (2003) used gravity darkening dependence of the light curve. Besides, Alonso \et (2004) presented different configurations that can produce signals resembling those produced by transiting planets, listing several strategies to recognize these false alarms. They suggested that multicolor photometry is an useful tool to recognize, for example, eclipsing binaries. Near-IR photometry can be also used to discard giant stars and stars with IR excess as well as the possibility to filter out larger companions observing host stars ellipsoidal variability (e.g., Sirko \& Paczynski 2003). Gallardo \et (2005) developed a method based on the well-calibrated surface brightness relation along with the correlation between mass and luminosity for Main-Sequence stars. This allows not only the classification of the most promising candidates, but also the measurement of the effective temperature and radius for the parent star.

As already mentioned, the OGLE project (Udalski 2002, 2003, 2004) has been an efficient photometric survey for discovering transiting extrasolar planets, proving that the photometric technique of detection of exoplanets can be successfully applied. During the years 2003 and 2005, the ``planetary and low-luminosity companion'' catalogues were published by the OGLE team. In these campaigns, several fields of the Galactic disk were monitored in search for low-depth transits, and 113 objects were discovered with transit depths shallower than 0.08 magnitudes in the $I$ band.

In this work, which is the continuation of the work published by Gallardo \et (2005), we present new optical and near-IR photometry and measure the
properties of a sample of 27 stars with observed transits (from the 113 just
mentioned), constraining astrophysical parameters of both
stellar and companion objects. We use the method developed by Gallardo \et (2005) and we
focus our analysis on the G, K and M-type parent stars. We select the most
interesting stellar objects to host extrasolar planets from the
transit light curve, transit depth and then derive stellar parameters
such as effective temperature, stellar radius and distance. 

The paper is
organized as follows: in \S2, we present our observations followed by
data reduction in \S3. In \S4, we briefly discuss the method we used to
determine both stellar and companions physical parameters, and the
inferred properties are presented in \S5. The conclusions given in \S6,
close the paper.

\section{Observations}

 In April and May 2005, we carried out several observations of OGLE extrasolar planetary transit candidates of its campaigns during the years 2003 and 2005. We observed every target for 2 or 3 hours (about 100 images for a single star). A total of 14 stars were observed for planetary transits in their light curves. We complemented these with observations (5 images per star) of other OGLE stars to obtain near-IR photometry, and made a near-IR photometric catalogue for the OGLE candidates.
We used the SoFI IR 
camera at the ESO New Technology Telescope (NTT) for our observations. The SoFI instrument is 
equipped with a HAWAII HgCdTe detector of 1024 $\times$ 1024 pixels, characterized by a 5.4 e$^{-}$/ADU gain, a readout noise of 2.1 ADU, and a dark
current of less than 0.1 e$^{-}$ s$^{-1}$. We observed in the Large
 Field mode, giving a 4.9 $\times$ 4.9 
arcmin$^{2}$ field. All measurements were made through the $J$, $H$ and $K_S$-band filters. 
%($\lambda_0 = 2{.}162
% \mu$m and $\Delta\lambda = 0{.}275 \mu$m).

Using a backup program, additional observations of some OGLE planetary transit candidates (taken from those previously observed in the near-IR bands) were carry out with the SMARTS 1.0 m telescope at
the Cerro Tololo International Observatory (CTIO) in the $V$
and $I$ optical bands. We included in this sub-sample one target already observed in Gallardo \et (2005): OGLE-TR-111 (a confirmed planet) and 5 eclipsing binary systems observed and analyzed by Pont \et (2005): OGLE-TR-72, OGLE-TR-78, OGLE-TR-105, OGLE-TR-106 and OGLE-TR-123, in order to compare their results with our conclusions (see discussion in \S 5).
The camera was 
the Y4KCam with STA 4064$\times$4064 CCD with 15-micron pixels
mounted in an LN$_2$ dewar, with a field of view 20$\times$20 arcmin$^{2}$
square, giving a scale of 0.289-arcsec/pixel. The data were obtained
during two nights on 2009, March 14-15. The fields were monitored both in the
$V$ and $I$ band filters with 300 and 180 seconds exposure time for
each image respectively.

\section{Data reduction}

\subsection{Near-IR photometry}

 The reductions were made using IRAF tasks\footnote{IRAF is distributed by the NOAO, operated by Universities for Research in Astronomy, Inc.}. 
 First, a cross-talk correction was applied, 
taking into account the detectors sensitivity 
difference between the upper and the lower half. 
Then the sky substraction was applied. The whole data
 set was acquired using ``dither'' mode (regular 
offset for a specific number of images). 
%In our case we used ``dither-5'' and ``dither-9''). 
These 
contiguous sky images were used to generate local sky background close in time for each of the offset images.
 Then, the appropriately scaled skies were substracted
 from the images. Finally, we applied flat-field 
corrections to all images and aligned them. For the
 flat-fields we used the correction images provided
 by the NTT SciOps team\footnote{See www.ls.eso.org/lasilla/sciops/ntt/sofi/reduction/flat\_fielding.html}, and the alignment was done 
with \verb+lintran+ and \verb+imshift+ task.

 The calibration of the near-IR photometry was made using
2MASS, with common stars between our fields and stars
 listed in 2MASS Point Sources Catalog. In this calibration we
 obtained zero point of 0.1 mag for the $K_S$-band 
photometry. Some stars used for the calibrations were checked
 against the Deep Near-Infrared Survey (DENIS) sources.

The average errors for the candidates are less than 0.1 mag. for the used near-IR filters. Finally, some candidates were not included in the analysis because they were classified as corrupted data.

\subsection{Optical photometry}

The reductions of the $V$ and $I$ band data were made using the IRAF {\textit{phot}} task. We have picked an
aperture size of 6 pixels and a dannulus of 2 pixels for substracting
the sky background. The transformation to the standard system was performed on the base of observations of three Landoldt's (1992) fields using the Bouger
linear relation by fitting the instrumental magnitudes for different
airmass observations. The target stars have $14.50\leq V\leq18.69$ mag and
$13.79\leq I\leq16.72$ mag with errors that should not exceed $0.05$ and $0.07$
mag, respectively.

\section{Determining the stellar parameters}

This is the second paper using the same analysis presented in Gallardo \et (2005), thus, here we briefly discuss the main equations developed in order to determine physical parameters of the observed objects. Two steps are necessary in order to characterize
each star. Using the limb darkened angular diameter of the star
($\theta_{\mathrm{LD}}$), we can infer intrinsic magnitudes via the following relations:
\begin{eqnarray}
\log \left( \frac{\theta_{\mathrm{LD}}}{\mathrm{mas}} \right)&=&0.5 + 0.1(V-K)_o-0.2 K_o,
\label{mas1}\\
\log \left( \frac{\theta_{\mathrm{LD}}}{\mathrm{mas}} \right)&=&0.5 + 0.2(I-K)_o-0.2 K_o,
\label{mas2}
\end{eqnarray}
where the index ``$o$'' refers to unreddened values and where the rms. dispersions are less than $1\%$ and $3.7\%$ respectively
(Kervella \et 2004) and the final estimate is the weighted mean of
these estimates. Then we can infer the effective temperature
through the relation:
\begin{equation}
\log T_{\mathrm{eff}}= 4.2-\left[1.2\log \theta_{\mathrm{LD}}+0.2K_o - 0.6 \right]^{1/2}.
\label{te}
\end{equation}
Finally, the combination of relations (\ref{mas1}), (\ref{mas2}) and
(\ref{te}) yields an estimate of \rstar~by its dependence on the
assumed distance as:
\begin{equation}
\left(\frac{R_*}{R_\odot}\right)= 4.4 \times 10^{10} \left( \frac{d}{\mathrm{kpc}} \right) 
 \tan \left( \frac{ \theta_{\mathrm{LD}} }{2} \right).
\label{ra}
\end{equation}

The second step comes from the properties of Main-Sequence stars: the strong correlation
between luminosity and mass ($L \sim M^{\beta}$), and between mass and radius ($M \sim R^{\alpha}$),
hence between effective temperature and radius of the form:
\beq
\left( \frac{ T_{\mathrm{eff}} }{T_{\mathrm{eff}\odot}} \right) \; = \; 
\left( \frac{ R_* }{R_{\odot}} \right)^{\left( \alpha \beta - 2 \right)/4} \; \approx \; 
\left( \frac{ R_* }{R_{\odot}} \right)^{0.64},
\label{eq:ms}
\eeq
using the standard values of $\alpha=1/0.8$, $\beta \approx 3.6$ and solar effective temperature 
$T_{\mathrm{eff}\odot} = 5770 $ K. In this way, if the distance exists,
both estimates agree and the parameters are fully consistent. On
the other hand, if there is no distance solution, then either the star is a
giant (for which equations (\ref{mas1}), (\ref{mas2}) and (\ref{te}) do not apply), or the assumed reddening at that distance is incorrect (necessary to calculate $(I-K)_o$). At last, the companion radius is derived using the standard relation from Seager \& Mallen-Ornelas (2003):
 
\beq
\left( \frac{R_{c}}{R_{*}} \right) \; = \; \left( 10^{\Delta I / 2.5} - 1 \right)^{1/2}, 
\label{eq:radius}
\eeq
where $\Delta I$ is the depth of the transit in magnitudes in the $I$ band, provided by OGLE. 
 
\section{Constrains on the stellar objects and companions}

Table \ref{tabla} lists the inferred stellar properties. In column (1) we give the name of the target, the derived distance and the effective temperature of the host star in column (2) and (3), respectively. The calculated companion radius and its error are shown in column (4) and (5), respectively. The stellar radius is lists in column (6) and the corresponding error is shown in column (7). The transit depth in the $I$ band is displays in column (8). Finally, column (9) presents the observed orbital period. 

\begin{figure}[h!]
\begin{center}
\includegraphics[angle=-90,scale=.3]{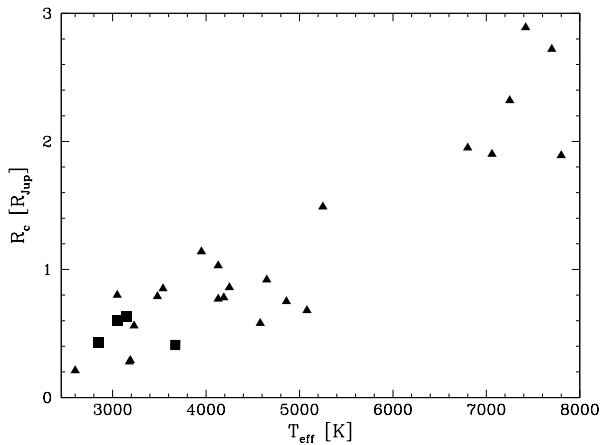}
\caption{Radius for the companions versus effective temperature of the host star for 27 observed OGLE objects. Error of $T_{\mathrm{eff}}$ is of $\pm$ 600 [K]. Filled squares indicate the promising candidates around M-type stars while the rest of the sample is indicated with filled triangles.}
\label{Rc-Teff}
\end{center}
\end{figure}

As this table shows, there are many M-type stars (Column (3)) in this sample. Analyzing these results, several companions
have radii comparable with the known close-in giant transiting
extrasolar planets and, even more interesting, some planetary companion
candidates are orbiting late M-type stars, as Fig.~\ref{Rc-Teff}
shows. This figure displays the effective temperature of the host star versus the derived transiting companion's radius. 

\begin{figure}[h!]
\begin{center}
\includegraphics[angle=0,scale=.3]{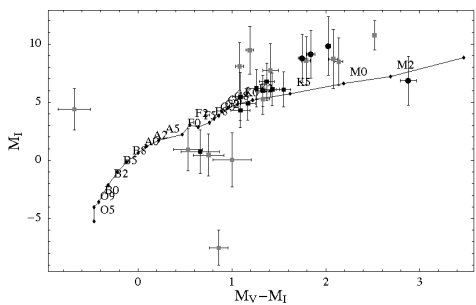}
\caption{Absolute color-magnitude diagram. Filled black circles denote the most promising M-type
  candidates. The filled black squares denote also promising candidates around F- and/or G-type stars. On the other hand, discarded candidates are represented by grey filled squares. Error bars are mainly due to distance
  uncertainties. The Main-Sequence stars loci have been plotted in solid black line.}
\label{Abs-Mag}
\end{center}
\end{figure}

We have made the following selection steps to find the most promising
candidates to host exoplanets. First, we made a spectral type and 
effective temperature analysis, using Table \ref{tabla} and
Fig.~\ref{Abs-Mag}, where we plot the $V-I$ and $I$ absolute values for the 27 stars in filled squares, along with the locus of the Main-Sequence stars, showed in solid line. Thus, we have selected the following M-type star candidates: OGLE-TR-61, OGLE-TR-74, OGLE-TR-123, OGLE-TR-143,
OGLE-TR-161, OGLE-TR-167, OGLE-TR-172 and OGLE-TR-173. Considering the obtained companion radii, in consistency with the median
radii of exoplanets found by transits, only the candidates OGLE-TR-61,
OGLE-TR-74, OGLE-TR-123, OGLE-TR-167 and OGLE-TR-173 are
the most promising stars to host extrasolar planets, and deserve
further spectroscopic observations. Very recently, Pietrukowicz \et (2010) conclude that the shape of the transit OGLE-TR-167 rules out its planetary nature based on VIMOS+VLT photometry. Pont \et (2005) presented an spectroscopic follow-up of OGLE-TR-123, but they claimed that more data is needed to confirm the planetary nature of the mentioned object. Their tentative result indicates that OGLE-TR-123 may be orbited by a brown dwarf or low-mass star near the hydrogen-burning limit, making OGLE-TR-123 an extremely interesting candidate. The {\textit{possible}} sub-stellar companion of the other four stellar objects (i.e., OGLE-TR-61, OGLE-TR-74, OGLE-TR-123 and OGLE-TR-173) orbit M-type stars.
 These are very interesting in their own right, because they appear to have small size stellar
companions like HD-149026b (Sato \et 2005), HAT-P-3b (Torres \et 2007)
or OGLE-TR-111b (Minniti \et 2007), with radius ranging from $\sim$ 0.5
\rjup~to 1.0 \rjup. Moreover we show the reliability of the method
proposed by Gallardo \et (2005), by confirming OGLE-TR-111 (Minniti \et
2007) as an exoplanet host star that has effective temperature 
and companion radius in total consistency with our
estimations.
 
%While for the exoplanet OGLE-TR-132b the estimations on
%radii are not well determined due to inconsistencies in distance
%estimation by the method mentioned, or probably in the $V$ and $K$
%obtained magnitudes. 
It is important to notice that some K-type stars that have companion radius well measured, should also be considered as good transit candidates. These are OGLE-TR-71, OGLE-TR-72, OGLE-TR-158 and OGLE-TR-164. One of these candidates, OGLE-TR-72 was however discarded following Pont \et (2005), who classified it as M eclipsing binary system.
 
Finally, using the same method described above, we select OGLE-TR-60, OGLE-TR-155, OGLE-TR-160, OGLE-TR-162 and OGLE-TR-163
as also good transit candidates to host exoplanets,
belonging to earlier F or G spectral type Main-Sequence stars.

On the other hand, we discard all of the rest for the
following reasons; OGLE-TR-75 presents substantial inconsistencies in
the magnitude for estimating accurate stellar properties; OGLE-TR-70, OGLE-TR-156 and OGLE-TR-159
are in disagreement when comparing effective
temperature inferred and spectral type classification; OGLE-TR-78,
OGLE-TR-105, OGLE-TR-106, OGLE-TR-142 and OGLE-TR-146
are either giants since the method based on the well-calibrated
surface brightness relation along with the correlation between mass
and luminosity for Main-Sequence stars, does not give acceptable
solutions, or their distances have been estimated with errors
that are far too high. It is important to note that candidates OGLE-TR-78, OGLE-TR-105 and OGLE-TR-106, were classified by Pont \et (2005) as M, G and M binary systems, respectively, making  our selection criteria result of these candidates a robust conclusion. Finally, candidates OGLE-TR-143, OGLE-TR-161 and
OGLE-TR-172 have low-mass companions whose radii are far too small to
be good transit candidates following our selection criteria.

\section{Conclusions}

We present the second paper, as a continuation of the results presented in Gallardo \et (2005), analyzing optical and near-IR photometry for candidate OGLE transits. We carry out a self-consistent characterization of systems identified by
the OGLE collaboration. By combining near-IR and optical observations, we obtained the
stellar parameters such as effective temperature, distance, mass and
radii of 27 transiting candidates using the method presented in Gallardo
\et (2005). This method provides accurate determination of the
aforementioned astrophysical properties, allowing us to further refine
the radii of the companions and make an efficient selection of possible
extrasolar planets.

Our main conclusions are: (1) The objects: OGLE-TR-70, OGLE-TR-72, OGLE-TR-75, OGLE-TR-78,
OGLE-TR-105, OGLE-TR-106, OGLE-TR-142, OGLE-TR-143, OGLE-TR-146, OGLE-TR-156, OGLE-TR-159, OGLE-TR-161, OGLE-TR-167 and OGLE-TR-172 can be
excluded from future observations; (2) We select M or K-type star
candidates OGLE-TR-61, OGLE-TR-71, OGLE-TR-74,
OGLE-TR-123, OGLE-TR-158, OGLE-TR-164 and OGLE-TR-173 as very good transit candidates to
host exoplanets; (3) We have also selected OGLE-TR-60, OGLE-TR-155, OGLE-TR-160, OGLE-TR-162 and OGLE-TR-163 as
some other possible Main-Sequence stars harboring planets. Dreizler \et (2007) selected OGLE-TR-138 (not included in our analysis), OGLE-TR-158,
OGLE-TR-160 and OGLE-TR-163 as possible stars harboring planets. We agree in the classification for 3 of their 4 best candidates. Moreover, our
estimate of the \teff~and $R_\mathrm{c}$ are also in agreement with their
work. This confirms the reliability of our present selection.
Finally, it is important to notice that candidates in group (2)
are all M or K spectral type stars potentially one of the first detections of
planetary companions orbiting a stellar object with this
characteristic. There is only one exoplanet transiting an M-type dwarf known: Glise 436b (Butler \et 2004; Gillon \et 2007). In this way, we mostly select candidates OGLE-TR-61,
OGLE-TR-74, OGLE-TR-123 and OGLE-TR-173 as the most
promising M-type stars to host extrasolar planets around, and plan to obtain
high-resolution follow-up observations to confirm our results by
determining the masses accurately and to conclude their nature.

%######################################
%#######################################

\begin{table*}
\caption{Physical properties for the 27 observed OGLE objects and their companions.} 
\begin{center}
\begin{tabular}{crrrrrrrr}\hline 
 \hline
\multicolumn{1}{c}{Name} &    
\multicolumn{1}{c}{$d^a$} &
\multicolumn{1}{c}{$T_{\mathrm{eff}}^b$}  & 
\multicolumn{1}{c}{$R_\mathrm{c}^c$} & 
\multicolumn{1}{c}{$\pm  R_\mathrm{c}$} &   
\multicolumn{1}{c}{$R_\ast$}&
\multicolumn{1}{c}{$\pm R_\ast$}  &
\multicolumn{1}{c}{$\Delta I$} &
\multicolumn{1}{c}{$P$} \\    
 \multicolumn{1}{c}{OGLE-TR-}       &  \multicolumn{1}{c}{[kpc]}       & \multicolumn{1}{c}{[K]}      &
\multicolumn{1}{c}{[$R_\mathrm{Jup}$]}  &  \multicolumn{1}{c}{[$R_\mathrm{Jup}$]}  &  \multicolumn{1}{c}{[$R_\odot$]}  &
  \multicolumn{1}{c}{[$R_\odot$]} &   \multicolumn{1}{c}{[mag]} & \multicolumn{1}{c}{[days]} \\ \hline 
60&  2.35&  7800&  1.89 &0.30& 1.60 &0.26& 0.016 &  2.31   \\
61&     0.25&  3050	&       0.60 &0.06& 0.37&0.04  &    0.030&       4.26\\
70   &    0.25 &   3050&	         0.80&0.08& 0.36& 0.04 &     0.053   &     8.04\\
71   &    0.80 &   4250	&  0.86 &0.06& 0.62&0.05 &  0.022   &  4.19 \\
72   &    0.65 &   3950 &  1.14  &0.09 &0.55 &0.04&    0.048   &     6.85\\
74   &    0.25 &   3150  & 0.63 &0.06& 0.39  &0.04&  0.030    &     1.58\\
75   &    1.55&   5250  &    1.49 & 0.22& 0.86& 0.13&     0.034   &     2.64\\
78& 2.40  &7250   & 2.32 &0.40 &1.43& 0.25& 0.03 &   5.32\\
105& 2.80  &6800   & 1.95 &0.54 &1.29&0.36& 0.026  &   3.06\\
106$^{*}$& 6.00  & 7060  & 1.90 &0.10 & -- &--& 0.022  &  2.54 \\
111$^{**}$& 0.80  & 4650  & 0.92 &0.06 &0.71 &0.05& 0.019 & 4.02  \\
123  &    0.50 &    3670    &      0.41 &0.03 &0.49& 0.04 &      0.008  &     1.80\\
142$^{*}$   &    3.80 &  7700&   2.72&0.1 &--&--  &    0.034  &  3.06\\
143   &   0.04 &    2600&   0.21 &0.02& 0.29& 0.03 &    0.006  &  3.34\\
146$^{*}$&  5.80 &7420   & 2.89  &0.1&--&-- & 0.043 &  2.94 \\
155  & 1.10 &  5080 & 0.68 &0.06&0.82 &0.08 &  0.008& 5.27\\
156  &    0.25 &    3480&0.79 &0.07&0.45 &	0.04&  0.034&3.58\\
158  &    0.70 &    4130	&  0.77 &0.06&0.59 &0.04&    0.019&	  6.38\\
159  &    0.45  &    3540&0.85	&0.08& 0.47 &0.04&  0.038&	  2.12\\
160  &    0.50  &    4580&0.58	&0.04&0.70  &0.05&  0.008&	  4.90\\
161  &    0.25  &    3180&0.28	&0.03 &0.39 &0.04&  0.006&	  2.74\\
162  &    0.75  &    4860&0.75	&0.05  &0.76 &0.05& 0.011&	  3.75\\
163  &    0.65  &   4130&1.03	&0.08  &0.59 &0.04& 0.034&	 0.94\\
164   &   0.75  &    4190&0.78	& 0.06 &0.61 &0.05& 0.019&2.68\\
167   &   0.25  &    3230 & 0.56 &0.05 &0.40 &0.04&  0.022& 5.26\\
172   &   0.20  &    3190  &  0.29 &0.03&0.40&0.04&    0.006&	1.79\\
173  &    0.10  &    2850  &  0.43 &0.05 &0.33& 0.04&   0.019&	2.60\\
\hline
\end{tabular}
\end{center}

\noindent $^a$ Errors on the distance correspond to
  $\pm$ 0.25 kpc on average

\noindent $^b$ Errors on the effective temperature correspond to $\pm$ 600 K 

\noindent $^{c}$ $R_\mathrm{Jup}=0.103$ \rsun 

\noindent $^{*}$ Presumable giant stars (see discussion in \S 5)

\noindent $^{**}$ Already confirmed planet

\label{tabla}
\end{table*}

\begin{table*}
\caption{Physical properties for the 27 observed OGLE objects and their companions.} 
\begin{center}
\begin{tabular}{crrrrrr}\hline 
 \hline
\multicolumn{1}{c}{Name} &    
\multicolumn{1}{c}{$a$} &
\multicolumn{1}{c}{$V$}  & 
\multicolumn{1}{c}{$I$} & 
\multicolumn{1}{c}{$J$} &   
\multicolumn{1}{c}{$H$}&
\multicolumn{1}{c}{$K$}  \\
 \multicolumn{1}{c}{OGLE-TR-}       &  \multicolumn{1}{c}{[mag]}       & \multicolumn{1}{c}{[mag]}      &
\multicolumn{1}{c}{[mag]}  &  \multicolumn{1}{c}{[mag]}  &  \multicolumn{1}{c}{[mag]}  &
  \multicolumn{1}{c}{[mag]} \\ \hline 
60&0.016&  15.25  & 14.59&  14.16&  14.07&  13.81  \\
61& 0.03&  18.17  &  16.33&  14.82&  14.27&  13.87\\   
70&0.053&  17.87   & 16.68&   15.7&  14.18&  13.93 \\  
71&0.022&  17.55  &  16.22&  15.61&  15.13&  14.54   \\
72&0.048&  17.78   & 16.41&  15.61&  15.13&  14.54   \\
74& 0.03&  17.74   & 15.99&  14.37&  13.95&  13.58   \\
75&0.034&  16.18   & 16.86&  15.96&  14.87&  14.84   \\
78&  0.03  & 16.24  & 15.71 & 14.93  &14.68 & 14.64\\
105&  0.026  &16.93   & 15.93 & 15.09 &14.72 &14.9 \\
106& 0.022   & 17.84  & 16.72 & 15.99 &16.04 &16.3 \\
%109& 0.008   &  15.9 & 15.15 &13.72  &14.26 & 13.45\\
111&   0.019 & 16.93  &15.6  & 14.7 &14.22 &14.11 \\
    123&0.008&  18.69   &  15.81&  14.64&  14.31&  14.21   \\
%    132&0.011&  16.74  &  15.88&  15.25&  14.86&  14.99   \\
   142&0.034&   16.22  & 15.32&  15.18&  14.86&   14.9   \\
   143&0.006&   16.31 &  13.79&   12.1&   11.1&  10.73   \\
   146&0.043&   17.79 &  16.43&  15.73&  15.41&  16.02   \\
   155&0.008&   16.77  & 15.68&  14.82&  14.44&  14.29   \\
   156&0.034&   16.39  & 15.31&  13.78&  13.15&  12.98   \\
   158&0.019&   17.56 &  16.01&  14.89&   14.3&  14.29   \\
   159&0.038&   17.84 &  16.43&  15.32&  14.04&  14.17   \\
   160&0.008&   15.53 &  14.44&  13.57&  13.18& 13   \\
   161&0.006&   17.59  & 15.80&   14.5&  13.76&  13.55   \\
   162&0.011&   16.17 &  15.00&  14.09&  13.81&  13.67   \\
   163&0.034&   17.16  & 15.90&  14.99&   14.4&  14.14   \\
   164&0.019&   17.66 &  16.23&  15.07&  14.54&  14.51   \\
   167&0.022&   17.85  & 15.71&  14.41&  13.68&  13.41   \\
   172&0.006&   17.49  & 15.41&  13.93&  13.18&  12.99   \\
   173&0.019&   16.96  & 14.93&  13.31&  12.52&  12.34   \\
\hline
\end{tabular}
\end{center}

\label{tablaphot}
\end{table*}

\begin{acknowledgements}
DM and PP are  supported by FONDAP Center for Astrophysics 15010003,
BASAL Center for Astrophysics and Associated Technologies PFB-06. DM is also 
supported by MIDEPLAN Milky Way Millennium Nucleus. PP was also supported by the
Foundation for Polish Science through program MISTRZ and the Polish Ministry of Science and Higher Education through the grant N N203 301335.
\end{acknowledgements}

\clearpage

\end{document}